\documentclass[12pt]{iopart}
\usepackage{graphicx}
\usepackage{latexsym}
\usepackage{amsbsy}
\usepackage{verbatim}
\def\vektor#1{{\boldsymbol{#1}}}

\begin{document}

\date{\today}
\title{Temperature dependent magnetization dynamics of magnetic nanoparticles}

\author{ A. Sukhov$^{1,2}$ and J. Berakdar$^2$}
\address{$^1$Max-Planck-Institut f\"ur Mikrostrukturphysik,
Weinberg 2, D-06120 Halle/Saale, Germany\\
$^2$Institut f\"ur Physik, Martin-Luther-Universit\"at
 Halle-Wittenberg, Heinrich-Damerow-Str. 4, 06120 Halle, Germany}
\begin{abstract}
Recent experimental and theoretical studies show that the
switching behavior of magnetic nanoparticles can  be well
controlled by external time-dependent magnetic fields.
 In this work, we inspect  theoretically  the influence of
 the temperature and the magnetic anisotropy  on the spin-dynamics and the switching properties
  of single domain magnetic nanoparticles (Stoner-particles).
 Our theoretical tools are the Landau-Lifshitz-Gilbert equation
 extended as to deal with finite temperatures within a Langevine
 framework.
Physical quantities of interest are
 the minimum field amplitudes required for switching  and the corresponding reversal times of
  the nanoparticle's magnetic moment. In particular, we contrast
  the cases of  static and time-dependent external fields and
  analyze the influence of damping  for a uniaxial  and a cubic
  anisotropy.
\end{abstract}
\maketitle
\section{Introduction}
In recent years, there has been a surge of research activities
focused on the spin dynamics and the switching behavior of
magnetic nanoparticles \cite{HiOu01}.  These studies are driven by
potential applications in mass-storage media and fast
magneto-electronic devices. In principle, various techniques are
currently available for controlling or reversing
 the magnetization of a nanoparticle. To name but a few,
the magnetization can be reversed by a short laser pulse
\cite{Vomi05}, a spin-polarized electric current
 \cite{Slon96, Berg96} or an alternating magnetic field
 \cite{Thir03,Sun06,SuWa06,Zhang06,Xu05,Mor07,Lee07,Nem07,Riv06}.
 Recently \cite{Sun06},
 it has been shown for a uniaxial anisotropy  that the utilization
 of a weak time-dependent magnetic
 field  achieves a magnetization reversal
  faster than in the case
 of a static magnetic field.
For this case  \cite{Sun06}, however, the influence of the
temperature and the different types of anisotropy on the various
dependencies of the reversal process have not been addressed.
These issues, which are the topic of this present work, are of
great importance since, e.g. thermal activation affects decisively
the stability of the magnetization, in particular when approaching
the superparamagnetic limit, which restricts the density of  data
storage \cite{chantrell}. Here we study the possibility of fast
switching at finite temperature with weak external fields. We
consider magnetic nanoparticles with an appropriate size as to
display a long-range magnetic order and to be in a single domain
remanent state (Stoner-particles). Uniaxial and cubic anisotropies
are considered and shown to decisively influence the switching
dynamics. Numerical results are presented and analyzed for
iron-platinum nanoparticles. In principle, the inclusion of finite
temperatures in spin-dynamics studies is well-established
 (cf. \cite{Brown63,Garcia,Usad06,DenPRB06,Den06,HiOu01} and
 references therein) and will be followed here
 by treating finite temperatures on the level of Langevine dynamics.
 %
For the analysis of switching behaviour the Stoner and Wohlfarth model
(SW) \cite{StWo48} is often employed. SW investigated  the
energetically metastable and stable position of the magnetization
of   a single domain particle with uniaxial anisotropy in the
presence of an external magnetic field. They showed that
 the minimum \emph{static} magnetic field (generally referred to as the Stoner-Wohlfarth (SW)
 field or limit)
needed to coherently reverse the magnetization is dependent on the
direction of the applied field with respect to the easy axis. This
dependence is described by the so-called
  Stoner-Wohlfarth astroid. The SW findings rely, however,
  on a static model at zero temperature.
  Application of a time-dependent magnetic field
   reduces the required  minimum switching field amplitude  below the SW limit \cite{Sun06}.
   It was, however, not yet clear how  finite temperatures will
   affect these findings. To clarify this point,
   we utilize an extension of the
   Landau-Lifshitz-Gilbert equation \cite{LaLi35}
  including finite temperatures on the level of Langevine dynamics
  \cite{Brown63,Garcia,Usad06}.
Our analysis shows the reversal time to be strongly dependent  on
the damping, the temperature and the type of anisotropy. These
dependencies are also exhibited to a lesser extent by the critical
reversal fields. The  paper is organized as follows:  next section
2 presents details of the numerical scheme and the notations
whereas section 3 shows numerical results and analysis for
Fe$_{50}$Pt$_{50}$
 and  Fe$_{70}$Pt$_{30}$ nanoparticles. We then conclude with a brief summary.
 %
%
%
\section{Theoretical model}
In what follows we focus on systems with large spins such that
their magnetic dynamics can be described by the classical motion
of a unit vector $\vektor{S}$ directed along  the particle's
magnetization $\vektor{\mu}$, i.e. $\vektor{S}=\vektor{\mu}/\mu_S$
and $\mu_S$ is the particle's magnetic moment at saturation. The energetics of
the system is given by
\begin{equation}
\displaystyle \mathcal{H}\,=\mathcal{H}_A + \mathcal{H}_F.
\label{hamil0}
\end{equation}
where $\mathcal{H}_A$ ($\mathcal{H}_F$) stands for the anisotropy
 (Zeeman energy) contribution.
Furthermore, the anisotropy contribution is expressed as
$\mathcal{H}_A=-D f(\vektor{S})$ with $D$ being the anisotropy
constant. Explicit form of  $f(\vektor{S})$ is provided below.
The magnetization dynamics, i.e. the equation of motion for $\vektor{S}$,
is governed by the Landau-Lifshitz-Gilbert (LLG) equation
\cite{LaLi35}
\begin{equation}
\frac{\partial {\vektor{S}}}{\partial t}
\,=\,-\,\frac{\gamma}{(1+\alpha^2)}{\vektor{S}}
 \times \Big[{\vektor{B}_{e}}(t)\,+\,\alpha \big ({\vektor{S}}
  \times {\vektor{B}_{e}}(t) \big) \Big].
\label{llg3}
\end{equation}
Here we introduced  the effective field ${\mathbf B}_{e}(t)
 = - 1/(\mu_S)\partial \mathcal {H} /\partial {\mathbf S}$
  which contains the external magnetic field and the maximum anisotropy field for the uniaxial anisotropy
$B_A = 2 D / \mu_S$.  $\gamma$ is the gyromagnetic ratio and
$\alpha$ is the  Gilbert damping parameter.
The temperature fluctuations will be
described  on the level of the Langevine dynamics \cite{Brown63}.
This means, a time-dependent
  thermal noise ${\boldsymbol \zeta}(t)$ adds to the effective field ${\mathbf
  B}_{e}(t)$ \cite{Brown63}.
  ${\boldsymbol \zeta}(t)$ is  a Gaussian distributed white noise with zero mean and vanishing time correlator
\begin{equation}
\quad \langle \zeta_{i}(t')\zeta_{j}(t) \rangle =
 \frac{2 \alpha k_B T}{ \mu_s \gamma}\, \delta_{i,j}\, \delta(t-t'). \label{corr4}
\end{equation}
$i, j$ are Cartesian components, $T$ is the temperature and $k_B$
is the Boltzmann constant. It is convenient to express the LLG in
the reduced units
\begin{equation} b = \frac {B_e}{ B_A}, \quad
\tau=\omega_a t,\quad \omega_a=\gamma B_A.
\label{eq:units}\end{equation}
The  LLG  equation reads then
\begin{equation}
\frac{\partial {\vektor{S}}}{\partial \tau}
=\,-\,\frac{1}{(1+\alpha^2)}{\vektor{S}} \times
\Big[{\vektor{b}}(\tau)\,+\,\alpha \big ({\vektor{S}} \times
{\vektor{b}}(\tau) \big) \Big], \label{redllg5}
\end{equation}
where the effective field is now given explicitly by
\begin{equation}
{\vektor{b}}(\tau) =  -\frac{1}{\mu_S B_A}\frac{\partial
\mathcal{H}}{\partial \vektor{S}}+{\vektor{\Theta}}(\tau)
\label{feld6}
\end{equation}
with
\begin{equation}
\quad \langle \Theta_{i}(\tau')\Theta_{j}(\tau) \rangle=
\epsilon\, \delta_{i,j} \delta(\tau-\tau');\: \: \epsilon\,=\frac{
2 \alpha k_B T}{ \mu_s B_A }. \label{dim_noise7}
\end{equation}
The  reduced units are independent of the damping parameter
$\alpha$.
 In the following sections we use extensively the
parameter
\begin{equation} q=\frac{ k_B T}{D}.
 \label{eq:qd}\end{equation}
$q$ is a measure for the thermal energy in terms of the
anisotropy energy. And $d=D/(\mu_S B_A)$ expresses the anisotropy constant  in
units of a maximum anisotropy energy for the uniaxial anisotropy and is always $1/2$.
The stochastic LLG equation (\ref{redllg5}) in reduced units  (\ref{eq:units}) is solved numerically using the
Heun method which converges in quadratic mean to the solution of the LLG equation when interpreted in the sense of
Stratonovich \cite{Garcia}. For each type of anisotropy we choose the time step $\Delta \tau$ to be one thousandth
part of the corresponding period of oscillations. The values of the time interval in not reduced units for uniaxial and
 cubic anisotropies are $\Delta t_{ua}=4.61\cdot 10^{-15}\,s$  and $\Delta t_{ca}=64.90\cdot 10^{-15}\,s$,
 respectively, providing us thus with correlation times on the femtosecond time scale. The reason for the choice
 of such small time intervals is given in \cite{Brown63}, where it is argued that the spectrum of thermal-agitation
 forces may be considered as white up to a frequency of order $k_B T/h$ with $h$ being the Planck constant.
 This value corresponds to $10^{-13}\,s$ for room temperature. The total scale of time is limited by a
 thousand of such periods. Hence, we deal with around one million iteration steps for a switching process.
 Details of realization of this numerical scheme could be found in references \cite{Schnake95, Now01, Garcia}.
 We note by passing, that attempts have been made to obtain, under certain limitations,
  analytical results for
  finite-temperature spin dynamics using the Fokker-Planck equation (cf. \cite{DenPRB06, Den06} and references therein). For the general case discussed here one has however to resort to fully numerical approaches. 
\section{Results and interpretations}
\begin{figure}[!t]
   \begin{center}
   \vspace{.35cm}
    \includegraphics[width=0.4\textwidth]{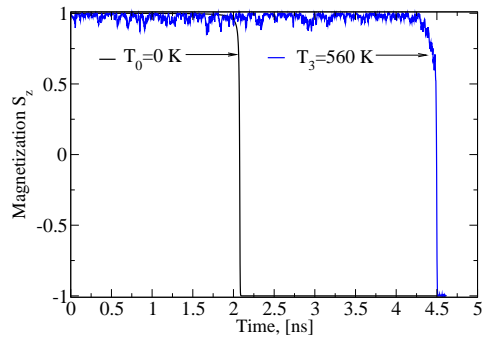}
    \caption{\label{fig_0} (Color online) Magnetization reversal of a nanoparticle
     when a static field is applied at zero Kelvin ($q_0=0$, black) and at reduced temperature $q_3=0.01\equiv 560 K$ (blue). The strengths of the fields in the dimensionless units
     (\ref{eq:units}) and (\ref{eq:qd}) are $b=1.01$ and $b=0.74$, respectively. The damping parameter is $\alpha=0.1$. The start position of the magnetization is given by the initial angle $\theta_0=\pi/360$ between the easy axis and the magnetization vector.}
   \end{center}
\end{figure}

We consider a magnetic nanoparticle  in a single domain remanent
state (Stoner-particle) with an effective anisotropy whose origin
can be magnetocrystalline, magnetoelastic and surface anisotropy.
We  assume the
 nanoparticle to have a spherical form, neglecting thus the shape anisotropy contributions.
 In the absence of external fields, thermal fluctuations may still drive the system out
 of equilibrium.
 Hence, the stability of the system as the temperature
 increases becomes an important issue.
  The time $t$ at which the magnetization of the system
  overcomes the energy barrier due to the thermal activation,
  also called the \textit{escape time}, is given by the Arrhenius law
\begin{equation}
t=t_{0}e^{\frac{D}{k_B T}}, \label{escape8}
\end{equation}
where the exponent is the ratio of the anisotropy to the thermal
energy.
 The coefficient $t_0$ may be inferred when  $D \gg k_B T$
 and for high damping  \cite{Brown63} (see  \cite{Klik90} for a critical
 discussion)
\begin{equation}
t_{0}=\frac{1+\alpha}{\alpha \gamma}\frac{\pi \mu_S}{2
D}\sqrt{\frac{k_B T}{D}}. \label{tzero9}
\end{equation}
Here we focus on two different types of
iron-platinum-nanoparticles:  The compound Fe$_{50}$Pt$_{50}$
which has a uniaxial anisotropy  \cite{Anto06,Osta03}, whereas the
system
 Fe$_{70}$Pt$_{30}$ possesses a  cubic anisotropy \cite{Anto05}.  Furthermore,
 the temperature dependence will be studied by
 varying $q$ (cf. eq.(\ref{eq:qd})).\\
For Fe$_{50}$Pt$_{50}$ the important parameters for simulations are the diameter of the nanoparticles $6.3\,nm$, the strength of the anisotropy $K_u=6\cdot 10^6 J/m^3$, the magnetic moment per particle $\mu_p=21518 \cdot \mu_B$ and the Curie-temperature $T_c=710 K$ \cite{Anto06, Osta03}.  The relation between $K_u$ and $D_u$ is $D_u=K_u V_u$, where $V_u$ is the volume of Fe$_{50}$Pt$_{50}$  nanoparticles. In the calculations for
Fe$_{50}$Pt$_{50}$  nanoparticles the following $q$ values were
chosen: $q_1=0.001$, $q_2=0.005$ or $q_3=0.01$ which correspond to
the real temperatures $56K$, $280K$ or $560K$, respectively (these
temperatures are
 below the blocking temperature). The
corresponding escape times are
 $t_{q_1}\approx 2\cdot 10^{217} s$, $t_{q_2}\approx  10^{75}s$ and
 $t_{q_3}\approx 7\cdot 10^{31}s$, respectively. In some cases we also show the results for an additional temperature $q_{01}=0.0001$ with the corresponding real temperature to be equal to $5K$. The corresponding escape time for this is $t_{q_{01}}\approx 10^{4300} s$.
  These times should be compared with the measurement period
  which is about $t_m\approx 5 \; ns$,  endorsing thus the
 stability  of the system during  the measurements.\\
For Fe$_{70}$Pt$_{30}$ the parameters are as follows: The diameter of the nanoparticles $2.3\,nm$, the strength of the anisotropy $K_c=8\cdot 10^5 J/m^3$, the magnetic moment per particle $\mu_p=2000 \cdot \mu_B$, the Curie-temperature is $T_c=420 K$ \cite{Anto05}, and 
 $D_c=K_c V_c$ ($V_c$ is the volume.)
For Fe$_{70}$Pt$_{30}$ nanoparticles the values of $q$ we choose
in the  simulations are $q_4=0.01$, $q_5=0.03$ or $q_6=0.06$ which
means that the temperature is  respectively   $0.3K$, $0.9K$ or
$1.9K$. The escape times are $t_{q_4}\approx  10^{34} s$,
$t_{q_5}\approx 2\cdot 10^{5}s$ and
 $t_{q_6}\approx 2 \cdot 10^{-2}s$, respectively. Here we also choose an intermediate value $q_{04}=0.001$ and the real temperature $0.03 K$ with the corresponding escape time to be equal to $t_{q_{04}}\approx 10^{430} s$. The measurement period is the same, namely about $5\, ns$. All values of the escape times were given for $\alpha=0.1$.\\
Central to this study are two issues:  The \emph{critical magnetic
field} and the corresponding \emph{reversal time}. The critical
magnetic field we define as the minimum field amplitude needed to
completely reverse the magnetization. The reversal time is the
corresponding time for this process. In contrast, in other studies
\cite{Sun06}  the reversal time is defined as the time needed for
the magnetization to switch from the initial position to the
position $S_z=0$, our reversal time is
 the time at which the magnetization reaches the very proximity of
  the antiparallel state (Fig. \ref{fig_0}). The difference in the definition is in so
  far important as
the magnetization position $S_z=0$ at finite temperatures
  is not stabile so it may switch back to
     the initial state due to thermal fluctuations and hence
      the target state is never reached.
%
\subsection{Nanoparticles having uniaxial anisotropy:
 {Fe}$_{50}$Pt$_{50}$ }
A Fe$_{50}$Pt$_{50}$ magnetic nanoparticle has a uniaxial
anisotropy whose direction defines  the $z$ direction. The
magnetization direction  $\vektor{S}$ is specified by   the
azimuthal angle $\phi$ and the polar angle $\theta$ with respect
to $z$.  In the presence of an external field $\vektor{b}$ applied
at an arbitrarily chosen direction, the energy of the system in
dimensionless units derives from
\begin{equation}
\displaystyle \tilde{\mathcal{H}}\,= -  d \cos^2 \theta - \vektor{S}\cdot \vektor{b}.
\label{u_a_hamil}
\end{equation}
The initial state of the magnetization is chosen to be close to $S_z = + 1$
and we aim at the target state $S_z = - 1$.
\subsubsection{Static field}
\begin{figure}[!t]
   \begin{center}
   \vspace{.35cm}
    \hspace{-.6cm}
    \includegraphics[width=0.28\textwidth,angle=-90]{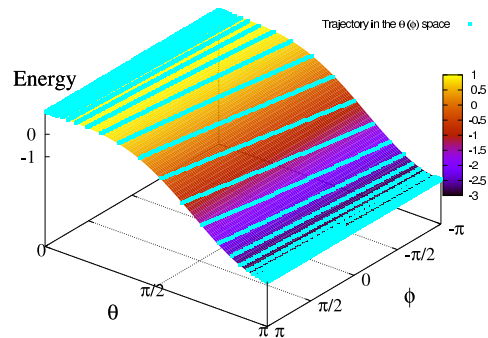}
    \caption{\label{fig_0b} (Color online) The
    trajectories of the magnetization unit vector parameterized by the angles $\theta$ and $\phi$
                                            at zero temperature. Other parameters are
                                           as in Fig. \ref{fig_0} for $q_0$.}
   \end{center}
\end{figure}

\begin{figure}[!t]
   \begin{center}
   \vspace{.35cm}
    \hspace{-.6cm}
    \includegraphics[width=0.4\textwidth]{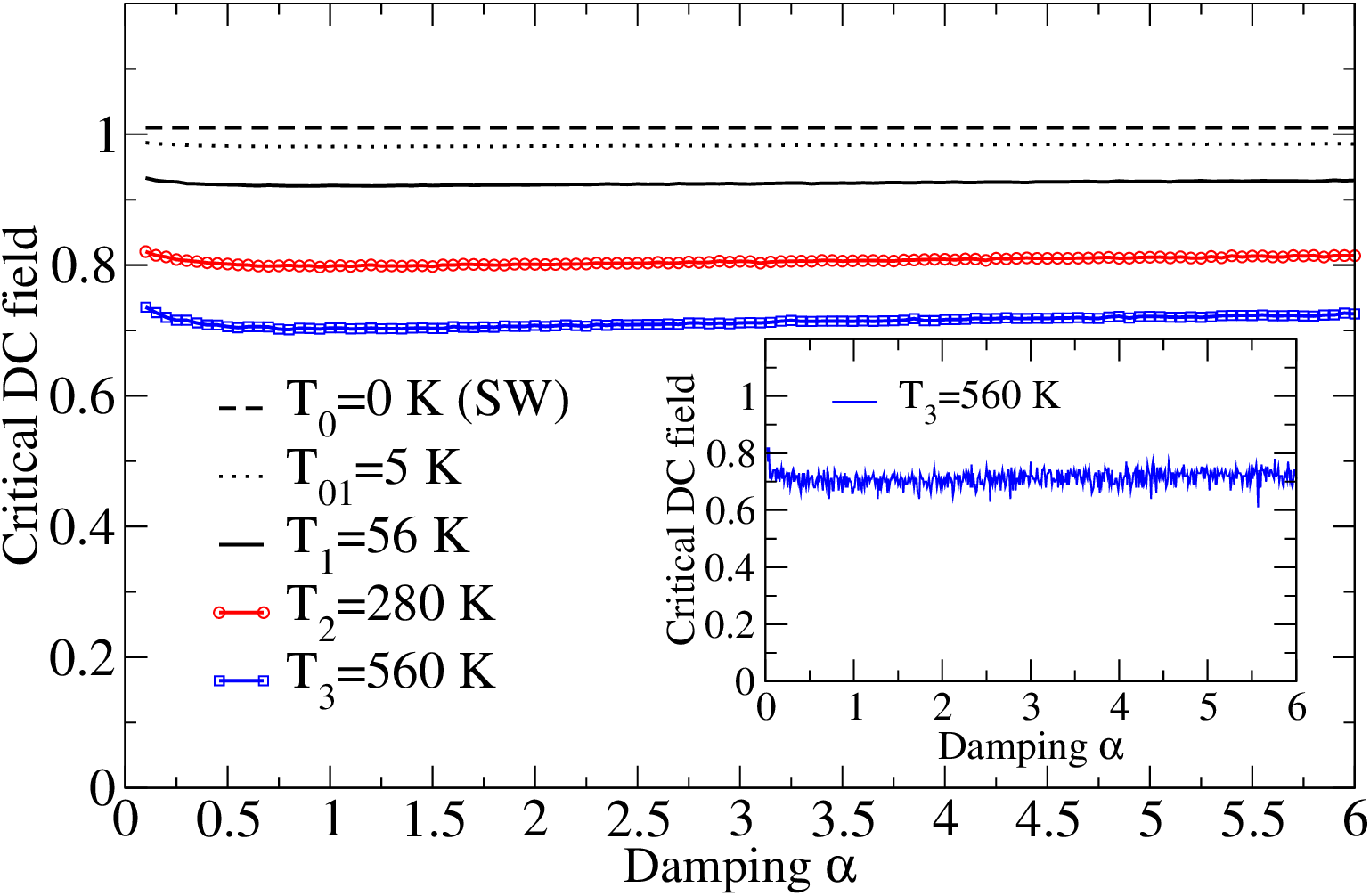}
    \caption{\label{fig_1} (Color online)
     Critical static field amplitudes vs. the damping parameters for different
      temperatures averaged over 500 times. Inset shows not averaged data for $q_3=0.01\equiv 560 K$.}
   \end{center}
\end{figure}

\begin{figure}[!t]
   \begin{center}
   \vspace{.35cm}
    \hspace{-.6cm}
    \includegraphics[width=0.4\textwidth]{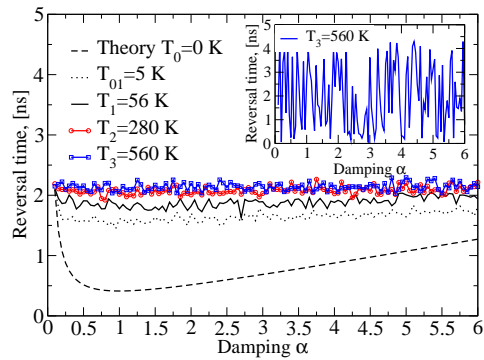}
    \caption{\label{fig_2} (Color online) Reversal times corresponding to the critical static fields in Fig. \ref{fig_1} vs. damping averaged over 500 cycles.
    Inset shows the as-calculated numerical results for $q_3=0.01\equiv 560 K$ (one cycle).}
   \end{center}
\end{figure}

For an external static magnetic field  applied antiparallel to the
$z$ direction  ($\vektor{b} = - b \vektor{e}_z$)
eq.(\ref{u_a_hamil}) becomes
\begin{equation}
\displaystyle \tilde{\mathcal{H}}\,= - d \cos^2 \theta +  b \cos \theta.
\label{u_a_hamil_phi_theta}
\end{equation}
To determine the critical field magnitude needed for the
magnetization reversal we proceed as follows (cf. Fig.
\ref{fig_0}): At first, the external field is increased in small
     steps. When the  magnetization reversal is achieved
     the corresponding values of the critical field versus the damping parameter
      $\alpha$ are plotted as shown in the inset of Fig. \ref{fig_1}.
      The  reversal times corresponding to the critical static field amplitudes of Fig. \ref{fig_1}
       are plotted versus damping  in Fig. \ref{fig_2}.\\
  In the Stoner-Wohlfarth (static) model  the mechanism
     of magnetization reversal is not due to damping.
      It is rather caused  by a  change of
      the energy profile in the presence of the field.
      The  curves displayed  on the energy surface in Fig. \ref{fig_0b}
       mark the magnetization motion in the $E$($\theta$,$\phi$) landscape.
    The magnetization
       initiates
 from $\phi_0=0$ and $\theta_0$ and ends up at $\theta=\pi$.
 As clearly can be seen from the figure,
 reversal is only possible if the initial state is energetically higher than the target state.
 This "low damping" reversal is, however, quite slow, which will be quantified
 more below. For the reversal at $T=0$, the SW-model predicts a minimum static
field strength, namely $b_{cr}=B/B_A=1$ (the dashed line in Fig.
\ref{fig_1} ).\\
  This  minimum field measured  with respect to the anisotropy field strength does
   not depend on the damping parameter $\alpha$, provided the measuring time is
    infinite. For $T>0$ the simulations were averaged over 500 cycles with the result shown in Fig. \ref{fig_1}.
The one-cycle data are shown  in the inset. Fig. \ref{fig_1}
evidences that
        with increasing
       temperature thermal fluctuations assist a weak magnetic field
       as to reverse the magnetization. Furthermore,
          the required critical field is increased  slightly at very large and
          strongly at very small damping with the minimum  critical field being at
          $\alpha\approx 1.0$.
             The reason for this  behavior is that for low damping  the second
             term of equation (\ref{llg3}) is much smaller than the first one,
             meaning that the system  exhibits a weak relaxation. In the absence
              of damping,  higher fields are necessary to switch
               the magnetization. For high $\alpha$, both terms in equation (\ref{llg3})
                become small (compared to a low-damping case) leading to a stiff
                magnetization and hence higher fields are needed
                to drive the magnetization.
                 For moderate damping,  we observe a minimum of
                   switching fields which is due to an optimal interplay  between  precessional
                    and  damping terms. Obviously, finite temperatures do not influence this general trend.\\
For the case of $q_0=0$, the Landau-Lifshitz-Gilbert equation of
motion can be solved analytically in spherical coordinates. The details of the solution can be found in Ref. \cite{Garcia} (eq. (A1)-(A8)). The final result of the solution in this reference differs, however, from the one given here due to to different geometries in these systems. In contrast to our alignment of the magnetization and the external field, the static field in Ref. \cite{Garcia} is applied parallel to the initial position of the magnetization. For
the solution, we assume that the magnetization
 starts at $\theta=\theta_0=\pi/360$ and arrives at $\theta=\pi$.
Note, that the expression
   $\theta\neq0$ is important only for zero Kelvin since the switching
    is not possible if the magnetization starts at $\theta_0=0$ (the vector product in
 equation (\ref{llg3}) vanishes).
     The reversal time in the SW-limit is then given by
\begin{equation}
t_{rev}=g(\theta_0,b)\frac{1+\alpha^2}{\alpha}, \label{trev10}
\end{equation}
where $g$ is defined as
\begin{equation}
g(\theta_0,b)=\frac{\mu_S}{2\gamma D}
\frac{1}{b^2-1}\ln\left(\frac{tg(\theta/2)^b\sin\theta}{b-\cos\theta}\right)\Big|^{\pi}_{\theta_0}
\label{eq:g}.\end{equation}
 From this relation we infer that  switching is possible only if
 the applied field is larger than the anisotropy field and the reversal time decreases with increasing $b$.
This conclusion is independent of the Stoner-Wohlfarth model and
follows directly
 from the solution of the LLG equation. An illustration is shown  by the dashed curve in Fig.
  \ref{fig_2}, which was a test to compare the appropriate numerical results with the analytical one. As our aim is the study of the  reversal-time dependence on the magnetic moment
and on the anisotropy constant, we deem the logarithmic dependence
in Eq.(\ref{eq:g}) to be weak  and write
\begin{equation}
g(b,\mu_S,D)\approx\frac{\mu_S}{\gamma}\frac{2D}{B^2\mu_S^2-4D^2}.
\label{trev11}
\end{equation}
This relation indicates that an increase in the magnetic moment
results in a decrease of the reversal time. The magnetic moment
enters in the Zeeman energy and therefore the increase in magnetic
moment is very similar to an increase in the magnetic field. An
increase of the reversal time with the increasing anisotropy
originates from the fact that the anisotropy constant determines
the height of the potential barrier. Hence, the higher
the barrier, the longer it takes for the magnetization to overcome it.\\
For the other temperatures the corresponding reversal times (also
averaged over 500 cycles) are shown in Fig. \ref{fig_2}.
     In contrast to the case $T=0$, where an appreciable dependence on damping is observed, the reversal times for finite temperatures show a weaker dependence on damping.
    If $\alpha\rightarrow0$ only the precessional motion of the magnetization
     is possible and therefore $t_{rev}\rightarrow\infty$. At high damping the system relaxes
     on a
      time scale that is  much shorter than the precession time, giving thus rise to an increase in
      switching times. Additionally, one can clearly observe the increase of the reversal times
      with increasing temperatures, even though these time remain on the nanoseconds time scale.
\subsubsection{Alternating field}
As was shown in Ref. \cite{Sun06, SuWa06, DenPRB06} theoretically and in
 Ref. \cite{Thir03} experimentally, a rotating alternating field with no
 static field being applied can also be used
 for the magnetization reversal. A circular polarized microwave field is applied
  perpendicularly
 to the anisotropy axis.  Thus, the Hamiltonian might be written in form of
  equation (\ref{u_a_hamil})
and the applied field is
\begin{equation}
  \displaystyle \vektor{b}(t) = b_0 \cos \omega t \vektor{e}_x + b_0 \sin \omega t \vektor{e}_y,
  \label{u_a_ac_field}
\end{equation}
where $b_0$ is the alternating field amplitude   and $\omega$ is
its frequency. For a switching of the magnetization the
appropriate frequency of the applied alternating field should be
chosen. In Ref. \cite{DenPRB06} analytically and in \cite {Sun06} numerically a detailed analysis of the optimal
frequency is given which is close to the precessional frequency of
the system. The role of temperature
and different types of anisotropy  have not yet been addressed, to our knowledge.\\
\begin{figure}[!t]
  \begin{center}
    \vspace{.35cm}
    \hspace{-.6cm}
    \includegraphics[width=0.4\textwidth]{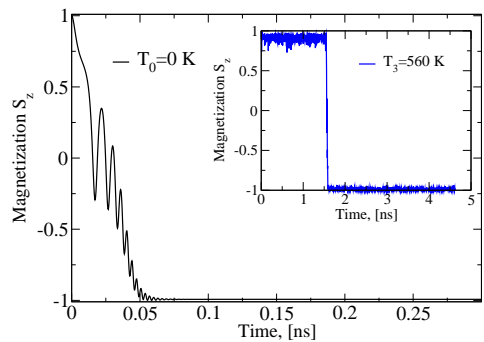}
    \caption{\label{fig_3} (Color online) Magnetization reversal
      in a nanoparticle using a time dependent field for $\alpha=0.1$ and at a zero temperature.
      The field strength and frequency  in the units (\ref{eq:units})
      are respectively
      $b_0=0.18$ and  $\omega=\omega_a/1.93$. Inset shows for this case the magnetization reversal for the temperature $q_3=0.01\equiv 560 K$ with $b_0=0.17$ and the same frequency.}
  \end{center}
\end{figure}
Fig. \ref{fig_3} shows  our calculations for the reversal process
at  two different temperatures. In contrast to the static case,
the reversal proceeds through  many oscillations  on a time scale
of approximately ten picoseconds. Increasing the temperature results in an
increase of the reversal time.

\begin{figure}[!t]
   \begin{center}
   \vspace{.35cm}
    \hspace{-.6cm}
    \includegraphics[width=0.28\textwidth,angle=-90]{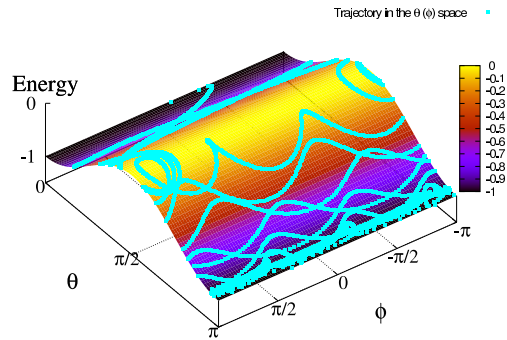}
    \caption{\label{fig_0c} (Color online) Trajectories followed
    by magnetization as specified by  $\theta$ and $\phi$
 for $q_0=0$.  Other parameters are $b_0=0.18$, $\alpha=0.1$ and  $\omega=\omega_a/1.93$. Energy-profile variations due
to the oscillating external field are not visible on this scale. }
   \end{center}
\end{figure}

Fig. \ref{fig_0c} shows the trajectory of the magnetization in the
E($\theta$,$\phi$) space related to the case of the alternating
field application. Compared with the situation  depicted in Fig.
\ref{fig_0b}, the trajectory reveals a quite delicate motion of
the magnetization. It is furthermore, noteworthy that the
alternating field amplitudes needed for the reversal (cf. Fig.
\ref{fig_4}) are substantially lower than their  static
counterpart, meaning that the energy profile of the system is not
completely altered by the external field.

Fig. \ref{fig_4} inspects the dependence of the minimum switching
field amplitude on damping. The critical fields are obtained upon
averaging over 500 cycles. The SW-limit lies by $1$ on this scale.
In contrast to the static case, the critical fields increase with
increasing  $\alpha$. In the low damping regime the critical field
is smaller than in the case of a static field. This behavior can
 be explained qualitatively by a resonant energy-absorption mechanism when the
frequencies of the applied field matches the frequency of the system. Obviously, at very low frequencies
(compared to the precessional frequency) the dynamics resembles the static case. \\
The  influence of the temperature on the minimum alternating field
amplitudes is depicted in Fig. \ref{fig_4}. With increasing
temperatures, the minimum amplitudes become smaller due to an
additional thermal energy pumped from the environment. The curves in this figure can be approached with two linear dependencies with different slopes for approximately $\alpha<1$ and for $\alpha>1$; for high damping it is linearly dependent on $\alpha$,
more specifically it can be shown that for high damping  the
critical fields behave as
\begin{equation}
  b_{cr}\approx\frac{1+\alpha^2}{\alpha}. \label{bac11}
\end{equation}
The proportionality coefficient contains the frequency of the
alternating field and the critical angle $\theta$. The solution
(\ref{bac11}) follows from the LLG equation solved for the case
when the phase of the external field follows temporally that of
the magnetization, which we checked numerically to be valid.

The reversal times associated with the critical  switching fields
are  shown in (Fig. \ref{fig_5}). Qualitatively, we observe the
same behavior as for the case of a static field. The values of the
reversal times for $T=0$ are, however, significantly smaller than for the
static case. For the same reason as in the static field case, an
increased temperature results in an increase of the switching
times.
\begin{figure}[!t]
  \begin{center}
    \vspace{.45cm}
    \hspace{-.6cm}
    \includegraphics[width=0.4\textwidth]{fig_7.eps}
    \caption{\label{fig_4} (Color online) Critical alternating field amplitudes vs.
    damping for different temperatures averaged over 500 times. Inset shows not
     averaged data for $q_3=0.01\equiv 560 K$.}
  \end{center}
\end{figure}
\begin{figure}[!t]
  \begin{center}
    \vspace{0.35cm}
    \hspace{-.6cm}
    \includegraphics[width=0.4\textwidth]{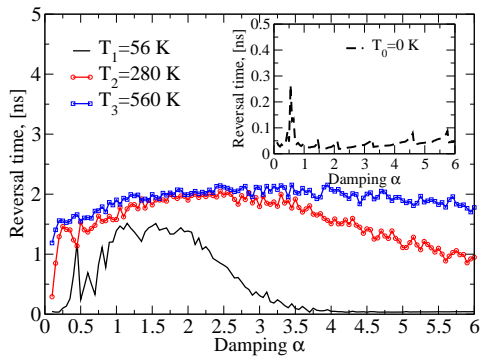}
    \caption{\label{fig_5} (Color online) The damping dependence
      of the reversal times corresponding to the critical field amplitudes of Fig. \ref{fig_4} for different temperatures.
       Inset shows the case of zero Kelvin.}
  \end{center}
\end{figure}

\subsection{Nanoparticles with cubic anisotropy: {Fe}$ _{70}$ Pt$ _{30}$}
Now we focus on another type of the anisotropy, namely a cubic
anisotropy which is supposed to be present for Fe$_{70}$Pt$_{30}$
nanoparticles \cite{Anto05}. The energetics of the system is then described by
the functional form
\begin{equation}
\displaystyle \tilde{\mathcal{H}}\,= - d (S_x^2 S_y^2 + S_y^2
S_z^2 + S_x^2 S_z^2) - \vektor{S}\cdot \vektor{b},
\label{c_a_hamil}
\end{equation}
or in spherical coordinates
\begin{equation}
\displaystyle \tilde{\mathcal{H}}\,= - d (\cos^2\phi \sin^2 \phi
\sin^4 \theta +\cos^2 \theta \sin^2 \theta) - \vektor{S}\cdot
\vektor{b}. \label{c_a_hamil_s_c}
\end{equation}
In contrast to the previous
section, there are more local minima or in other words more stable
states of the magnetization in the energy profile for the Fe$_{70}$Pt$_{30}$ nanoparticles. It can be shown that the minimum
barrier that has  to be overcome is $d/12$ which is twelve times
smaller than that in the case of a uniaxial anisotropy.
 The maximal one is only $d/3$.\\
The magnetization of these nanoparticles is first relaxed to the
initial state close to $\phi_0 = \pi/4$ and $\theta_0 =
\arccos(1/\sqrt{3})$, whereas in the target state it is aligned
antiparallel to the initial one, i. e. $\phi_e = 3\pi/4$
and $\theta_e = \pi - \arccos(1/\sqrt{3})$. In order to be
close to the starting state for the uniaxial
anisotropy case  we choose
 $\phi_0=0.2499 \cdot \pi$, $\theta_0=0.3042 \cdot \pi$.
\subsubsection{Static driving field}
\begin{figure}[!t]
   \begin{center}
   \vspace{.35cm}
    \hspace{-.6cm}
    \includegraphics[width=0.28\textwidth,angle=-90]{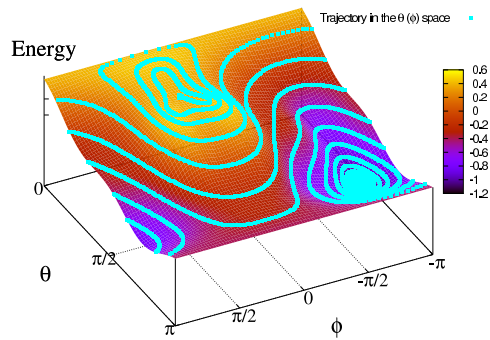}
    \caption{\label{fig_0e} (Color online) Trajectories of the magnetization  in the $\theta$($\phi$)
space ($q_0=0$). In the units (\ref{eq:units}) we
choose $ b=0.82$ and $\alpha=0.1$. }
   \end{center}
\end{figure}

\begin{figure}[!t]
   \begin{center}
   \vspace{.35cm}
    \includegraphics[width=0.4\textwidth]{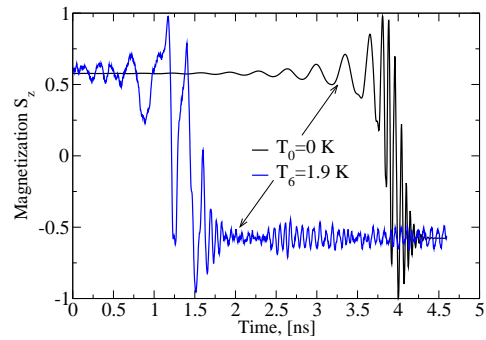}
    \caption{\label{fig_6} (Color online) Magnetization reversal of a nanoparticle
     when  a static field $b=0.82$ is applied and for $\alpha=0.1$ at zero temperature (black). The magnetization reversal for $\alpha=0.1$, $b=0.22$ and $q_6=0.06\equiv 1.9 K$ is shown with blue color.}
   \end{center}
\end{figure}

A static field is applied  antiparallel  to the initial state of
the magnetization, i.e.
\begin{equation}
  \displaystyle \vektor{b} = - b / \sqrt{3} (\vektor{e}_x + \vektor{e}_y + \vektor{e}_z).
  \label{c_a_field}
\end{equation}
In Fig. \ref{fig_0e} the trajectory of the magnetization in case
of an applied static field is shown. Similar to the previous
section the energy of the initial state lies higher than that of
the target state. The magnetization rolls down the energy
landscape to eventually end up by the  target state. The
trajectory the magnetization follows is completely different from the one for the
uniaxial anisotropy. Fig. \ref{fig_6} supplements this scenario of the
magnetization reversal by showing the time evolution of the $S_z$
vector. Because of the different anisotropy type, the trajectory
is markedly different from the case of the uniaxial anisotropy and
a static field. Here we show only the $S_z$ magnetization
component even though the other
components also have to be taken into account in order to avoid a wrong target state.\\
The procedure to determine the critical field amplitudes is
similar to that described in the previous section. In Fig.
\ref{fig_7} the critical fields versus the damping parameter for
different temperatures are shown. For $q_0$, the critical field
strength is  smaller than $1$. This is consistent insofar as the
maximum effective field for a cubic anisotropy is $\frac 2 3 B_A$.
In principle, the critical field turns out to be constant for all
$\alpha$ but for an infinitely large measuring time. Since we set
this time to  be about $5$ nanoseconds,  the critical fields increase
for small and high damping. On the other hand, at lower temperatures smaller
critical fields are sufficient for the (thermal
activation-assisted) reversal process.\\
 \begin{figure}[!t]
   \begin{center}
   \vspace{.35cm}
    \hspace{-.6cm}
    \includegraphics[width=0.4\textwidth]{fig_11.eps}
    \caption{\label{fig_7} (Color online)
     Critical static field amplitudes vs. the damping parameters for different temperatures averaged over 500 times.
     Inset shows not averaged data for $q_6=0.06\equiv 1.9 K$.}
   \end{center}
\end{figure}
 \begin{figure}[!t]
   \begin{center}
   \vspace{.35cm}
    \hspace{-.6cm}
    \includegraphics[width=0.4\textwidth]{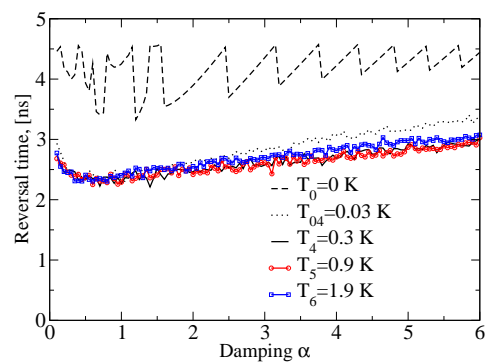}
    \caption{\label{fig_8} (Color online) Reversal times corresponding to the critical static fields of Fig. \ref{fig_7} vs. damping averaged over 500 times.}
   \end{center}
\end{figure}
The behaviour of the corresponding switching times presented in Fig. \ref{fig_8}
 only supplements the fact of too low measuring time,
 which is chosen as $5\,ns$ for a better comparison of
 these results with ones for uniaxial anisotropy. Indeed, constant jumps
 in the reversal times for $T=0\,K$ as a function of damping can be observed.
 The reason why the reversal times for finite temperatures are lower is as follows:
  The initial state for $T=0\,K$ is chosen to be very close to equilibrium.
  This does not happen for finite temperatures, where the system due to
   thermal activation jumps out of equilibrium (cf. see Fig. \ref{fig_6}).
\subsubsection{Time-dependent external field}
\begin{figure}[!t]
   \begin{center}
   \vspace{.35cm}
    \hspace{-.6cm}
    \includegraphics[width=0.28\textwidth,angle=-90]{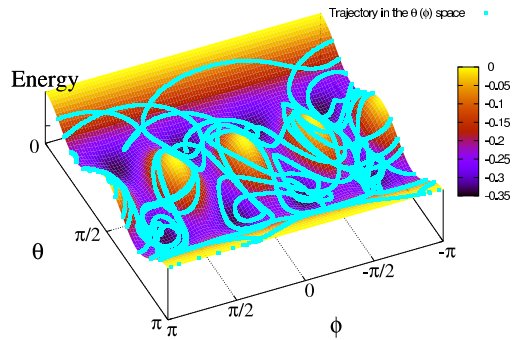}
    \caption{\label{fig_0f} (Color online) Trajectories of the magnetization vector specified by
    the angles  $\theta$ and $\phi$
at zero temperature. The chosen parameters are $b_0=0.055$ and  $\omega=\tilde{\omega_a}/1.93$, where $\tilde{\omega_a}=2/3\omega_a$. }
   \end{center}
\end{figure}

\begin{figure}[!t]
\begin{center}
    \vspace{.35cm}
    \hspace{-.6cm}
    \includegraphics[width=0.4\textwidth]{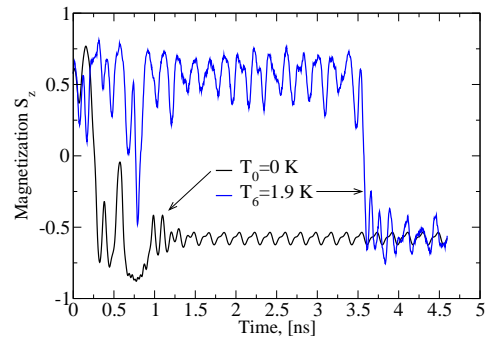}
    \caption{\label{fig_9} (Color online)
    Magnetization reversal in a nanoparticle using
     a time dependent field for $\alpha=0.1$ and $q_0$ (black) and for $q_6=0.06\equiv 1.9 K$ (blue).
     Other parameters are as in Fig. \ref{fig_0f}.}
  \end{center}
\end{figure}

\begin{figure}[!t]
  \begin{center}
    \vspace{.45cm}
    \hspace{-.6cm}
    \includegraphics[width=0.4\textwidth]{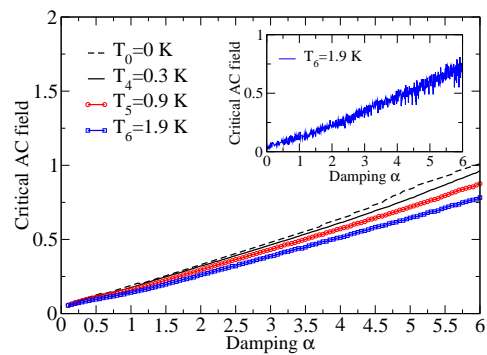}
    \caption{\label{fig_10} (Color online) Critical alternating field amplitudes vs.
     damping for different temperatures averaged over 500 cycles. Inset shows the single cycle data at $q_6=0.06\equiv 1.9 K$.}
  \end{center}
\end{figure}

\begin{figure}[!t]
  \begin{center}
    \vspace{0.35cm}
    \hspace{-.6cm}
    \includegraphics[width=0.4\textwidth]{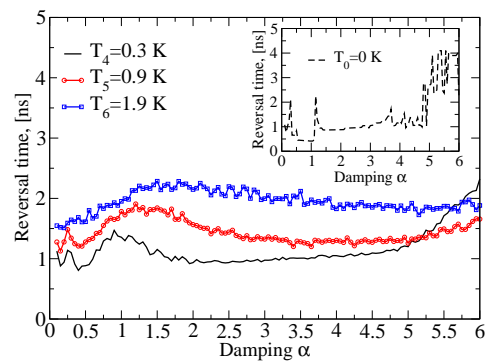}
    \caption{\label{fig_11} (Color online) The damping dependence
      of the reversal times corresponding to the critical fields of the Fig. \ref{fig_10} for different temperatures averaged over 500 runs.
      Inset shows the $T=0$ case.}
  \end{center}
\end{figure}

Here we consider the case of an alternating field that rotates in
the plain perpendicularly to the initial state of the
magnetization. It is possible to switch the magnetization with a
field rotating in the $xy-$ plane but the field amplitudes  turn
out to be larger than those when the field rotates perpendicular
to the initial state. For the energy this means that  the field
entering equation (\ref{c_a_hamil_s_c}) reads
\mathindent2pt
\begin{eqnarray}
 \vektor{b}(t) &=& (b_0 \cos \omega_1 t \cos \phi_0 + b_0 \sin \omega_1 t \sin \phi_0 \cos \theta_0)
  \vektor{e}_x \nonumber\\&& + (- b_0 \cos \omega_1 t \sin \phi_0 + b_0
                               \sin \omega_1 t \cos \phi_0 \cos \theta_0) \vektor{e}_y +
                               (- b_0 \sin \theta_0 \sin \omega_1 t) \vektor{e}_z,
\label{c_a_field_ac}
\end{eqnarray}
where $b_0$ is the alternating field  amplitude  and $\omega_1$ is
the frequency associated with the field. This expression is
derived upon a rotation of the field plane by
the angles $\phi_0 = \pi / 4$ and $\theta_0 = \arccos(1/\sqrt{3})$.\\
The magnetization trajectories depicted in  Fig. \ref{fig_0f}
reveal two interesting features: Firstly, particularly for small
damping, the energy profile changes very slightly (due to the
smallness of $b_0$) while energy is pumped into the system during
many cycles. Secondly,
  the system switches mostly in the vicinity of  local minima to acquire  eventually   the
  target state. Fig. \ref{fig_9} hints on the  complex character of the magnetization dynamics in this case.
As in the static field case with  a cubic anisotropy the critical
field amplitudes shown in Fig. \ref{fig_10} are smaller than those
 for a uniaxial anisotropy. Obviously, the reason is that the potential barrier associated with this anisotropy
 is smaller in this case, giving rise to smaller amplitudes.
 As before an increase in temperature leads
 to a decrease in the critical fields.\\
The reversal times shown in Fig. \ref{fig_11} exhibit the same
 feature as in the cases for uniaxial anisotropy: With increasing temperatures the
 corresponding reversal times increase. A physically  convincing  explanation of the (numerically stable)
 oscillations for the
 reversal times is still outstanding.
\section{Summary}

In this work we studied the critical field amplitudes required for
the magnetization switching of Stoner nanoparticles and derived
the corresponding reversal times for static and alternating fields
for two different types of anisotropies. The general trends for
all  examples discussed here can be summarized as follows:
 Firstly,  increasing the temperature results in a decrease of all
  critical fields regardless of the anisotropy type.
  Anisotropy effects decline with increasing temperatures making it easier to switch the magnetization.
  Secondly,
   elevating the temperature increases the corresponding reversal times.
    Thirdly,  the same trends are observed for different temperatures:
    The critical field amplitudes for
    a static field depend  only slightly  on $\alpha$, whereas the critical alternating field amplitudes
    exhibit  a  pronounced dependence on damping.
In the case of a uniaxial anisotropy we find  the critical
alternating field amplitudes to be smaller than those for a static
field, especially in the low damping regime and for finite
temperatures. Compared with a static field, alternating fields
lead to smaller switching times ({\bf $T=0\,K$}).
 However, this is  not the case for the cubic anisotropy.
The markedly different trajectories for the two kinds of
anisotropies endorse the qualitatively different magnetization
dynamics. In particular, one may see that for a cubic anisotropy
and for an alternating field the magnetization reversal takes
place through the local minima leading to smaller amplitudes of
the applied field. Generally, a cubic anisotropy is smaller than
the uniaxial one giving rise to smaller slope of critical fields,
i.e. smaller alternating field amplitudes.\\
It is useful to contrast our results with those of Ref.
\cite{DenPRB06}. Our  reversal times for AC-fields increase with
increasing temperatures. This is not in  contradiction with the
findings of \cite{DenPRB06} insofar as we calculate the switching
fields at first, and then deduce the corresponding reversal times.
If the switching fields are kept constant while increasing the
temperature \cite{DenPRB06} the corresponding reversal times
decrease. We note here that  experimentally known values of the damping parameter are, to our knowledge, not larger than $0.2$. The reason why we  go beyond this value is twofold. Firstly, the values of damping are only well known for thin ferromagnetic films and it is not clear how to extend them to magnetic nanoparticles. For instance, in FMR experiments   damping values  are obtained from the widths of the corresponding curves of absorption. The curves for nanoparticles can be broader due to randomly oriented easy anisotropy axes and, hence, the values of damping could be larger than they actually are. Secondly, due to a very strong dependence of the critical AC-fields (Fig. \ref{fig_4},  e.g.) they can even be larger than static field amplitudes. This makes the time-dependent field disadvantageous for switching in an extreme high damping regime.\\
Finally, as can be seen from all simulations, the corresponding
reversal times are much more sensitive a quantity than their
critical fields. This follows from the expression (\ref{trev10}),
where a slight change in the magnetic field $b$ leads to a sizable
 difference  in the reversal time. This circumstance is the basis
 for our choice to  average all the reversal times and fields over many times.
This is also desirable in view of an experimental realization,
for example, in FMR experiments or using a SQUID technique  quantities like
critical fields and their reversal times are averaged over
thousands of times.
The results presented  in this paper are of relevance to the heat-assisted magnetic recording, e.g. using a laser source. Our calculations do not specify the source of thermal excitations but they capture the
spin dynamics and switching behaviour of the system upon thermal excitations. 
%
\section*{Acknowledgments}
This work is supported by the International Max-Planck Research
School for Science and Technology of Nanostructures.

\end{document}